**Using the Data Agreement Criterion to Rank Experts' Beliefs**

Duco Veen[1*], Diederick Stoel[2], Naomi Schalken[1], Rens van de Schoot[1,3]

[1] Department of Methods and Statistics, Utrecht University, Utrecht, Netherlands
[2] ProfitWise International, Amsterdam, Netherlands
[3] Optentia research Focus Area, North-West University, Potchefstroom, South-Africa
[*] Corresponding author, email: d.veen@uu.nl

## Abstract

Experts' beliefs embody a present state of knowledge. It is desirable to take this knowledge into account when doing analyses or making decisions. Yet ranking experts based on the merit of their beliefs is a difficult task. In this paper we show how experts can be ranked based on their knowledge and their level of (un)certainty. By letting experts specify their knowledge in the form of a probability distribution we can assess how accurately they can predict new data, and how appropriate their level of (un)certainty is. The expert's specified probability distribution can be seen as a prior in a Bayesian statistical setting. By extending an existing prior-data conflict measure to evaluate multiple priors, i.e. experts' beliefs, we can compare experts with each other and the data to evaluate their appropriateness. Using this method new research questions can be asked and answered, for instance: Which expert predicts the new data best? Is there agreement between my experts and the data? Which experts' representation is more valid or useful? Can we reach convergence between expert judgement and data? We provided an empirical example ranking (regional) directors of a large financial institution based on their predictions of turnover.

## USING THE DATA AGREEMENT CRITERION TO RANK EXPERTS'

## BELIEFS

In the process of scientific inference, the knowledge and beliefs of experts can provide vital information. Experts' beliefs represent the current state of knowledge. It is desirable to be able to include this information in analyses or decision making processes. This can be done by using the Bayesian statistical framework. In Bayesian statistics there are two sources of information, prior knowledge and data (Gelman, Stern, & Rubin, 2014; Lynch, 2007; Zyphur, Oswald, & Rupp, 2015). The prior can be composed of expert knowledge (Bolsinova, 2016; O'Hagan et al., 2006; Zondervan-Zwijnenburg, van de Schoot-Hubeek, Lek, Hoijtink, & van de Schoot, 2017). Yet deciding which expert yield the most appropriate information remains a critical challenge, for which we present a solution in this paper.

To be able to include expert knowledge in Bayesian statistics, it must be represented in the form of a probability distribution. This can be done via a process called expert elicitation. Elicitation entails the extraction of expert knowledge and translating this knowledge into the probabilistic representation (O'Hagan et al., 2006). By using a probabilistic representation we include both knowledge and (un)certainty of experts. However, experts are forced to use the representation system that belongs to the statistical realm. Therefore it is essential that the elicitation process is carefully constructed so we do not introduce unnecessary and unjust bias.

Once expert knowledge is elicited and data is collected, it is desirable to find a measure that naturally compares two pieces of information. The measure should assess the extent to which the data and expert knowledge resemble and conflict with each other. As the expert knowledge can be contained within a prior it seems logical to assess the discrepancy or similarity of such a prior with respect to the data by means of a prior-data conflict measure.

Several prior-data conflict measures have been developed, for instance Box (1980) was the first to propose the use of the prior predictive distribution and test if the collected data was unlikely for this predictive distribution. The predictive distribution is the totality of all samples that could occur if the model assumptions were true. By reference of the density of the observed data to the predictive reference distribution, it can be evaluated if the data are unlikely to be generated by the model. A modification to this approach was proposed by Evans and Moshonov (2006) who suggested the restriction of the method to minimal sufficient statistics. For critics of this method see Evans and Jang (2011).

Both the methods of Box (1980) and Evans and Moshonov (2006) result in a p-value, and thus leave the determination of the existence of prior-data conflict up to debate depending on an arbitrary cut-off value. Further criticism comes from Young and Pettit (1996) who argue that a measure being based on a tail area does not produce the required behavior, if two priors are both specified at the correct value the more precise prior is not always preferred. A desirable property for a prior-data conflict measure would be to measure how one probability distribution diverges from a second expected probability distribution, not the distance between distributions. The Data Agreement Criterion (DAC) proposed by Bousquet (2008) is based on Kullback-Leibler (KL) divergences which ensures that prior-data conflict decisions are not only based on the tail area. Furthermore the DAC incorporates a clear classification of prior-data conflict.

Prior-data conflict measures are currently used to evaluate for example, the suitability of certain priors in the estimation of models or to uncover potential suitability problems with design, prior or both. Examples can be found in for instance Walley, Smith, Gale, and Woodward (2015); Fu, Celeux, Bousquet, and Couplet (2015) and Fu, Couplet, and Bousquet (2015). We found no previous use of prior-data conflict measures to rank sources of prior information. Yet when we have two experts some very interesting questions can already be answered, for instance: Which expert predicts the new data best? Is there agreement between my experts and the data? Which expert's representation is more valid or useful? Can we reach convergence between expert judgement and data?

Predicting new data can be an indication of knowledge of the factors underlying the data generating mechanism. Thus if one expert outperforms another in predicting new data this might be indicative of more expertise. Concerning the question if we can reach convergence between expert judgement and data it is interesting to consider the difference between aleatory and epistemic uncertainty. Aleatory uncertainty is uncertainty due to randomness or chance, e.g. market volatility, whilst epistemic uncertainty is uncertainty due to a lack of knowledge. By expressing expert knowledge and data in the same framework a learning process can start that has the potential to eliminate epistemic uncertainty and reveal the true extend of the aleatory uncertainty. This also implies, for the question which representation is more useful, that being overconfident can be a bad thing. When all epistemic uncertainty is eliminated there is still an appropriate amount of uncertainty to be specified, the extend of the aleatory uncertainty.

In the remainder of this paper we present the following work. We provide a detailed description of the DAC and show why this measure is especially suitable to compare expert judgement and data. As the DAC currently determines the degree of prior-data conflict of one prior we propose an adjustment of the statistic in this paper to allow the ranking of multiple sources of prior information, i.e. multiple experts' beliefs. Finally we provide an empirical example to show that the adapted DAC can be used to compare and rank several experts based on their beliefs. In the empirical example we rank experts from a large financial institution based on their predictions of new data concerning turnover. The empirical study in this article received approval from our internal Ethics Committee of the Faculty of Social and Behavioural Sciences of Utrecht University. The letter of approval can be found in the data archive for this study along with all other code and data, as far as contracts permit us, to ensure everything presented in this paper is reproducible. The data archive can be found on the Open Science Framework (OSF) webpage for this project as https://osf.io/u57qs.

## Expert-Data (Dis)Agreement

Within this section we provide a detailed and mathematical description of the DAC before proposing the adaptation that allows the ranking of multiple experts' beliefs at the same time. The DAC is based on a ratio of KL divergences therefore we will first describe the KL divergence (Kullback & Leibler, 1951).

### Kullback-Leibler divergence

The KL divergence describes measurements of informative regret, or in other words it measures the loss of information that occurs if one choses a certain distribution even though another is the preferred distribution. This loss of information or informative regret is expressed in a numerical value and the higher this value is the more loss of information is present, i.e. the greater the discrepancy is between the two distributions. The KL divergence is calculated by the following function

$$KL(\pi_1||\pi_2) = \int_{\Theta} \pi_1(\theta) \, log \frac{\pi_1(\theta)}{\pi_2(\theta)} \, d\theta, \qquad (1)$$

where $\Theta$ is the set of all accessible values for the parameter $\theta$, $\pi_1(\theta)$ denotes the preferred distribution and $\pi_2(\theta)$ denotes the chosen distribution that approximates the preferred distribution. In Figure 1 it can be seen what KL divergences between two normal distributions look like. The value of the KL divergence is equal to the area under the curve of the function. The greater the discrepancy between the distributions, the greater the area under the curve. This also follows from Equation 1, if the two distributions are equal than $\pi_1(\theta)$ / $\pi_2(\theta)$ equals one everywhere. As $log(1) = 0$ there is no longer any density and there is thus no area under the curve and the KL divergence, or loss of information is equal to zero. To support understanding of the KL divergence we build a shiny application that provides an interactive variant of Figure 1 which can be found via the OSF webpage at https://osf.io/u57qs.

If we are able to represent both the data and the expert knowledge in a distributional form, a discrepancy between the two can be expressed by the KL divergence between the two. As we might have multiple experts but only one source of data it seems natural that the data be considered the reference distribution which is approximated by the experts' beliefs expressed as probability distributions. That this is indeed the case in the DAC we will see in the following where we elaborate on the details of this prior-data conflict measure developed by Bousquet (2008).

**Data Agreement Criterion**

The DAC, as mentioned before, is a ratio of two KL divergences. A KL divergence provides an indication of the discrepancy between two distributions, yet has no inherent decision on when a certain amount of loss of information relevant. To be able to objectively conclude when prior-data conflict exists, the DAC compares the loss of information that a certain prior has with respect to the data with the loss of information that a benchmark prior has with respect to the data. The KL divergence between the chosen prior and the data is the numerator in the ratio whilst the KL divergence between the benchmark prior and the data is the denominator in the ratio. A benchmark prior should be chosen such that the posterior distribution is completely dominated by the observed data (Bernardo, 1979). In other words the benchmark prior should be an uninformative prior. A benchmark prior can be for instance a Jeffreys prior (Jeffreys, 1946), however Bousquet (2008) mentions that in applied studies a convenient or intuitive prior measure seems reasonable. Using the notation of Bousquet, the DAC can be expressed by

$$\text{DAC} = \frac{KL[\pi^J(\theta|\mathbf{y})||\pi(\theta)]}{KL[\pi^J(\theta|\mathbf{y})||\pi^J(\theta)]} \tag{2}$$

where $\pi(\theta)$ denotes prior that is to be evaluated, e.g. the expert's beliefs, $\pi^J(\theta)$ denotes a benchmark prior and $\pi^J(\theta|\mathbf{y})$ denotes the posterior distribution, derived from the benchmark prior and the observed data $\mathbf{y}$.

$\pi^J(\theta|\mathbf{y})$ can be considered as a fictitious expert that is perfectly in agreement with the data, having no prior knowledge and being informed by the observations (Bousquet, 2008). It is therefore the ideal reference distribution in comparison to which we measure the loss of information. The benchmark, being an uninformative prior, should by definition not be conflicting with the data and therefore serves as a good reference point. If a prior conflicts less with the data than the benchmark does, we should not consider the prior to be in prior-data conflict. If a prior conflicts more with the data than the benchmark prior does, we do consider the prior to be in prior-data conflict. Hence if the DAC > 1, we conclude prior-data conflict, because the KL divergence of the prior is larger than the KL divergence of the benchmark prior, otherwise we conclude no prior-data conflict.

To illustrate the calculation of the DAC we provide an numerical example together with a visual representation that can be found in Figure 2, note that the colors of the distributions in Figure 2 correspond to the colors in the text. Consider the case in which $\pi^J(\theta|\mathbf{y}) \sim N(0,1)$, $\pi(\theta) \sim N(0.5,1)$ and $\pi^J(\theta) \sim N(0,900)$. The DAC is than calculated by taking the ratio of the following two KL divergences, $KL[\pi^J(\theta|\mathbf{y})||\pi(\theta)] = 0.125$ and $KL[\pi^J(\theta|\mathbf{y})||\pi^J(\theta)] = 2.902$, such that DAC=0.125/2.902=0.043. The DAC<1, thus we conclude no prior data conflict exists, $\pi(\theta)$ is a better approximation of $\pi^J(\theta|\mathbf{y})$ than $\pi^J(\theta)$.

**Extension to multiple experts**

The DAC such as described in the section above determines prior-data conflict for a single prior that is to be evaluated. However, when we have multiple experts that each hold their own beliefs and we express each of these in the form of a probability distribution we can ask some interesting questions. In Figure 3 we see some examples of situations that we could encounter. In panel A of Figure 3 we see a situation in which experts differ in their predictions and their (un)certainty. The question that arises from the situation in panel A is which of these predictions best approximates the information that the data provides us? Panel B of Figure 3 shows a scenario in which the experts are predicting similar to each other but all differ with respect to the data. The question that arises from the situation in panel B is which of the two is correct, the data or the experts?

To be able to answer these types of questions we need to extend the DAC to incorporate multiple experts' priors, which are to be evaluated against the same posterior distribution, reflecting the data, and the same benchmark prior. The DAC thus needs to become a vector of length D resulting in

$$\text{DAC}_d = \frac{KL[\pi^J(\theta|\mathbf{y})||\pi_d(\theta)]}{KL[\pi^J(\theta|\mathbf{y})||\pi^J(\theta)]} \qquad (3)$$

where the subscript *d* denotes the different input for *D* experts so $\text{DAC}_d = \text{DAC}_1, \dots, \text{DAC}_D$ and $\pi_d(\theta) = \pi_1(\theta), \dots, \pi_D(\theta)$. This extension of the KL divergence in which not one but a vector of models are entered to be compared with the preferred model is straightforward and has previously been described in the context of the Akaike Information Criterion (AIC; Akaike, 1973), e.g. Burnham and Anderson (2002, chapter 2).

**Influence benchmark**

The choice for a specific benchmark can influence the results of the $\text{DAC}_d$. Bousquet (2008) mentions that in applied studies a convenient or intuitive prior for the benchmark seems reasonable. However it is important to realize that the choice for a benchmark prior does influence the results of the analysis in the sense that the cut-off value for determining prior-data conflict will shift as the KL divergence between $\pi^J(\theta|\mathbf{y})$ and $\pi^J(\theta)$ differs. Yet as long as the benchmark prior is an uninformative prior in the sense that the posterior distribution is dominated by the data, $\pi^J(\theta|\mathbf{y})$ will remain largely unchanged. This ensures that the $DAC_d$ has the nice property that when multiple experts are compared their ranking does not change dependent on which uninformative benchmark is chosen. This follows from the stability of $\pi^J(\theta|\mathbf{y})$ which ensures that the KL divergences between $\pi^J(\theta|\mathbf{y})$ and $\pi_d(\theta)$ are stable.

Concerning the benchmark it is useful to note that the benchmark need not be restricted to an uninformative prior, but using an informative prior changes the interpretation and behavior of the DAC. When $\pi^J(\theta)$ is informative $\pi^J(\theta|\mathbf{y})$ is sensitive to the specification of $\pi^J(\theta)$ and the KL divergence between $\pi^J(\theta|\mathbf{y})$ and $\pi_d(\theta)$ need no longer be stable, influencing the ranking of the experts. The choice for $\pi^J(\theta)$, if it is informative, will influence the cut-off value for determining prior-data conflict will shift.

To show the above described behavior visually we present the results of a simulation study in Figure 4. We show four different conditions, that is four different choices for benchmark priors, to illustrate the change in behavior for the $DAC_d$. In all four situations we use the same data, $\mathbf{y}$, which is a sample of 100 from a standard normal distribution. $\pi_d(\theta)$ is always a normal distribution and we show the $DAC_d$ values for normal distributions with a means running from the mean of the data -4 to the mean of the data +4 and standard deviations running from 0.1 to 3. The four panels show different conditions for the benchmarks such that in panel A $\pi^J(\theta) \sim N(0,10000)$, in panel B $\pi^J(\theta) \sim N(0,1)$, in panel C $\pi^J(\theta) \sim U(-50,50)$ and in panel D $\pi^J(\theta) \sim N(5,0.5)$. It can be seen that for the two uninformative priors in panels A and C the behavior of the $DAC_d$ is stable. We would expect to draw the same conclusions and rank experts in the same way independent of the choice of either benchmark. Yet when we specify an informative benchmark such as in panels B and D we see that both the behavior of the $DAC_d$ and the determination of prior-data conflict shift. In panel B an informative and accurate benchmark leads almost invariable to concluding prior-data conflict for $\pi_d(\theta)$. In panel D the informative but inaccurate benchmark leads us to conclude prior-data conflict only if $\pi_d(\theta)$ is in the wrong location and has a very small variance.

The simulation study presented in Figure 4 shows that the choice for a certain benchmark can influence your results, so even if a convenient or intuitive prior seems reasonable it should be carefully chosen. Researchers should be aware that their ranking is stable as long as an uninformative prior is chosen but it might not be if the benchmark prior contains information.

## Empirical Example

To show that the $DAC_d$ can be used to evaluate and rank several experts based on their beliefs we conducted an empirical study. The team that participated consisted of 11 experts, 10 regional directors and one director. All were eligible to be included in the study. Seven experts were randomly invited to participate in the research, if any of the selected experts did not want to participate they were classified as not selected in the research. In this way we avoided the possibility of group pressure to participate. In the end four out of the seven selected experts participated in an elicitation. The experts ($D = 4$) provided forecasts concerning average turnover per professional in the first quarter of the year 2016. The (regional) directors are considered experts in knowledge concerning market opportunities, market dynamics and estimating the capabilities of the professionals to seize opportunities. Based on these skills we expected that they could predict the average turnover per professional in the entire country in the first quarter of 2016. All information related to the empirical study can be found on the OSF webpage for this paper at https://osf.io/u57qs.

### Elicitation procedure

To get the experts to express their beliefs in the form of a probability distribution we make use of the Five-Step Method (Veen, Stoel, Zondervan-Zwijnenburg & van de Schoot, 2017). To encapsulate the beliefs of the expert, the Five-Step Method actively separates two elements of the knowledge of the expert, the tacit knowledge of the expert and their (un)certainty. In step one a location parameter is elicited from the expert. This location parameter captures the tacit knowledge of the expert. To verify that the representation of the beliefs is accurate, step two is the incorporation of feedback implemented through the use of elicitation software. Experts can accept the representation of their beliefs or adjust their input. In step three the (un)certainty of the experts is obtained and represented in the form of a scale and shape parameter. Step four is to provide feedback using elicitation software to verify the accurate representation of the expert's (un)certainty, which they can either accept or they can adjust their input until the representation is correct. The fifth step is to use the elicited expert's beliefs, in this case to determine their DAC score.

The experts first performed a practice elicitation for their own sales team before moving on to the whole country. The practice run enabled them to acquaint themselves with the elicitation procedure and software we used. Only in the case that the director participated this practice run would not be possible. The elicited distributions where restricted to be skewed normal distributions such that $\pi_d(\theta) \sim SN(\mu_0, \sigma_0^2, \gamma_0)$, where subscript $d$ denotes expert $d=1,\ldots,D$, $\mu_0$ denotes the prior mean, $\sigma_0^2$ denotes the prior variance and $\gamma_0$ denotes the prior skewness. The shape parameter $\gamma_0$ is based upon a general method for the transformation of symmetric distributions into skewed distributions as described by Equation 1 in Fernandez and Steel (1998). Table 1 provides an overview of the elicited distributions for the four experts in this empirical study. The distributions are based upon transformed data to avoid revealing business-sensitive information.

**Ranking the experts**

The predictions of the experts concerned the average turnover per professional ($N$=104). The benchmark we have chosen is a uniform distribution running from 0 to 5 which, given the scale of the transformed data and the impossibility of obtaining a negative turnover, seems an intuitive and uninformative prior. This is in line with the prior used by Bousquet (2008) in his Example 1 concerning a normal model. We obtained the posterior distribution using the `rjags` R-package (Plummer, 2016), such that $\pi^J(\theta|\mathbf{y}) \sim N(\mu_1, \sigma_1^2)$ where $\mu_1$ denotes the posterior mean and $\sigma_1^2$ denotes the posterior variance. We used 4 chains of 25,000 samples after a burn-in period of 1000 samples per chain. Visually inspection and Gelman-Rubin diagnostics did not point towards problems with convergence of the chains and inspection of the autocorrelation plots showed no issues concerning autocorrelation. Table 2 displays the KL divergences for $\pi_d(\theta)$ and $\pi^J(\theta)$ with $\pi^J(\theta|\mathbf{y}) \sim N(2.29, 0.01)$ as preferred distribution and the resulting $DAC_d$ scores and ranking. Figure 5 visually presents all relevant distributions concerning the empirical study. Figure 6 panels A through E visually presents all KL divergences from Table 2.

The results show that expert four provided the best prediction out of the experts. Experts one and two provided similar predictions concerning their tacit knowledge, they expected almost the same value for the location parameter, however expert one was less certain about this prediction. As the prediction of the location was not entirely correct, the increased uncertainty of expert one means this expert provided more plausibility to the regions of the parameter space that were also supported by the data. Therefore expert one is rewarded with a better DAC score in comparison to expert two. In this case the differences even cause a different conclusion, namely expert one is not in prior-data conflict and expert two is in prior-data conflict. Expert three provided a prediction that to a large extend did not support the same parameter space as the data. In fact expert three provides a lot of support for regions of the parameter space that the data did not support. The discrepancy between expert three and the data was of such proportions that besides expert two also for expert three we concluded a prior-data conflict to exist. If we would have had no information beforehand, except knowing the region within which the average turnover per professional could fall, we would have lost less information than by considering the predictions of experts two and three. If decisions should be made concerning average turnover per professional, decision makers would be wise to consult expert four, as this expert seems to have the best knowledge of the underlying factors driving these results.

## Discussion

The use of KL divergences raises two important methodological issues, see Burnham and Anderson (2002, chapter 2) for an elaborated discussion. First, the preferred model should be known. Second the parameters should be known for the model that is evaluated, i.e. the formalized expert prior. The issues make the KL divergence a measure that according to some, see for instance Burnham and Anderson, cannot be used for real world problems and previously led to the development of the AIC (Akaike, 1973), which uses the relative expected KL divergence. The AIC deals with the two issues by taking the preferred model as a constant in comparing multiple models and using the maximum likelihood estimates for the parameters of the models to be evaluated, introducing a penalty term for the bias this induces.

We conclude that we can use the KL divergence in the context of the $DAC_d$ and with the following reasoning. We define $\pi^J(\theta|\mathbf{y})$ to be the preferred distribution as it reflects a fictional expert that is completely informed by the data and thus it is known. In the case of the empirical example the data is even the true state of affairs, these where the actual realizations of the turnover for each professional. Concerning that the parameter for the models to be evaluated, $\pi_d(\theta)$, should reflect the exact beliefs of the experts. We use the Five-Step Method (Veen et al., 2017) which incorporates feedback at each stage of the elicitation, ensuring that experts confirm that their believes are accurately represented by the location, shape and scale parameters. We acknowledge that the parameters exactly representing an expert's beliefs cannot be exactly known, but we feel confident that the procedure we use at least aims to obtain very accurate representations. As experts can continue to adapt their input until they are satisfied with the representation of their beliefs, this should overcome problems with the second issue.

While we use $\pi^J(\theta|\mathbf{y})$, and thus know the truly preferred distribution, and we firmly believe that we properly represent the experts' beliefs, it seems highly implausible that a DAC score of 0 can be attained. Predicting future events entails incorporating an estimate for both epistemic and aleatory uncertainty. Yet even if epistemic uncertainty is eradicated it is unlikely one estimates precisely the optimal location and exactly the optimal amount of aleatory uncertainty.

Although we do not explicitly separate aleatory and epistemic uncertainty within the procedure, we are able to gain an indication of the appropriate amount of aleatory uncertainty. $\pi^J(\theta|\mathbf{y})$ provides an excellent indication of the appropriate aleatory uncertainty. Given that one had no knowledge beforehand and is rationally guided by the data, following probabilistic reasoning, one arrives at the posterior belief represented by $\pi^J(\theta|\mathbf{y})$. The posterior described the range of values that would have been plausible given this information. Given that the posterior is constructed via probabilistic reasoning, the uncertainty concerning the parameter is an indication of appropriate aleatory uncertainty given the information one has. This indication is however conditional on the fact that the data provide an accurate representation of the state of affairs.

Given that we can attain information on both the expected value for the parameter of interest, the appropriate amount of aleatory uncertainty and the quality of the approximation by each expert, we can start a learning process. By sharing the reasons behind the choices they made, experts can learn from one another as the evidence shows which reasoning lead to the most accurate predictions. The data can inform the experts so they can adjust their estimates and uncertainty. Through this evaluation expertise can increase and in the long run convergence should be reached between both different experts' predictions and between the experts and the data. When this convergence in reached, this indicates that, at least part of, the epistemic uncertainty is gone and we have better understanding of the data generating processes and are better able to make informed decision.

In the empirical example we can already see some opportunities for learning. For example, expert three misestimated the location of the parameter, which indicates at least to some extend faulty or missing tacit knowledge. By starting a dialogue with the other experts, he or she could learn why they all estimated the average turnover per professional to be higher. Expert one and two had almost identical predictions concerning the location but expert one expressed more uncertainty. Perhaps this indicated more acknowledgement of epistemic uncertainty, a dialogue could shed more light on the differences in choices of expert one and two.

Concerning the appropriateness of the ranking that is obtained using the $DAC_d$, we have the following to add. One could argue that perhaps the sample entails extreme data. Yet even if this is true, the experts should have considered the data to be plausible, for it did occur. Thus if an expert exhibits prior-data conflict or does not perform well compared to others, this expert simply did not expect that these data where likely or plausible. By incorporating (un)certainty in the evaluation, the $DAC_d$ produces the required behavior to fairly compare experts' beliefs. As noted earlier, Young and Pettit (1996) argue that a prior-data conflict measure should not be based on a tail area as a more precise prior would not always be preferred if two priors are both specified at the correct value. We argue that this property would in fact also be undesirable in a prior-data conflict measure. Given that it is appropriate to take aleatory uncertainty into account, a prior can be over specific such that it does not do right by the aleatory principles underlying the data generating mechanism. The $DAC_d$ rewards specifying an appropriate amount of uncertainty and penalizes overconfidence. We believe that the properties of the $DAC_d$ are such that, given unbiased and unpolluted data, the $DAC_d$ ranking of experts truly insinuates differences in expertise.

**Author contributions**

DV, DS and RvdS mainly contributed to the study design. DV and NS programmed and verified the statistical analyses for the $DAC_d$. DV programmed the elicitation software. The elicitations have been facilitated by DV and DS. DV wrote and revised the paper with feedback and input from DS NS and RvdS. RvdS supervised the project.

**Funding**

The project was supported by a grant from the Netherlands Organization for Scientific Research: NWO-VIDI-452-14-006.

**Conflict of interest statement**

The authors declare that the research was conducted in the absence of any commercial or financial relationships that could be construed as potential conflict of interest.

**Acknowledgements**

We are grateful to all participants of the empirical study for their time, energy and predictions. Also we would like to thank the company for allowing us access to their resources and information thereby enabling us to provide empirical support for the theoretical work.

*Figure 1*. KL divergences between two normal distributions. In this example $\pi_1$ is a standard normal distribution and $\pi_2$ is a normal distribution with a mean of 1 and a variance of 1. The value of the KL divergence is equal to the area under the curve of the function.

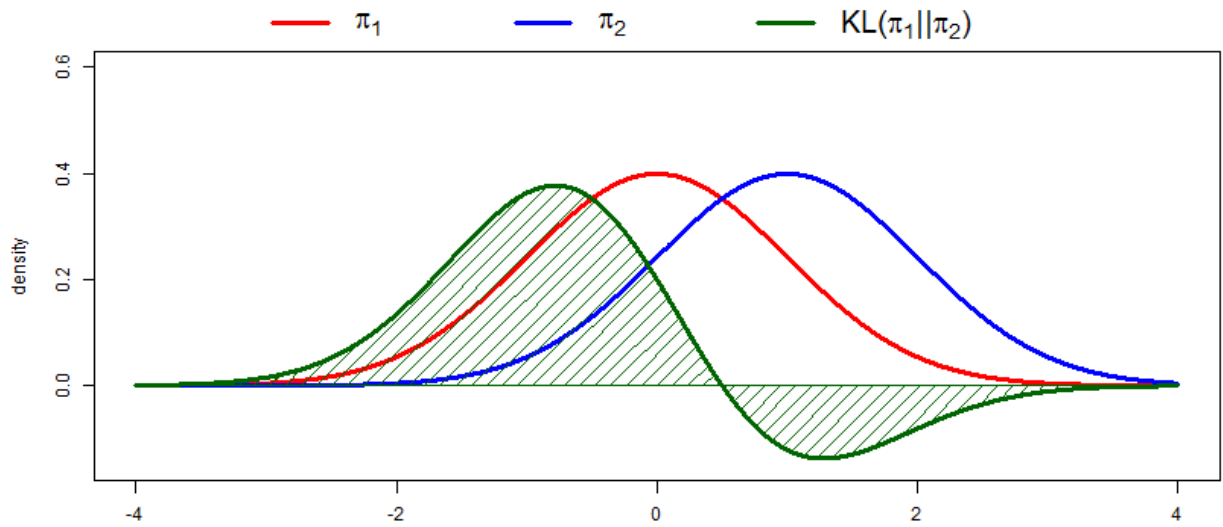

*Figure 2*. Calculating the DAC. In this example $\pi^J(\theta|\mathbf{y})$ is a standard normal distribution, $\pi(\theta)$ is a normal distribution with a mean of 0.5 and a variance of 1 and $\pi^J(\theta)$ is a normal distribution with a mean of 0 and a variance of 900. The DAC < 1, thus no prior-data conflict is concluded.

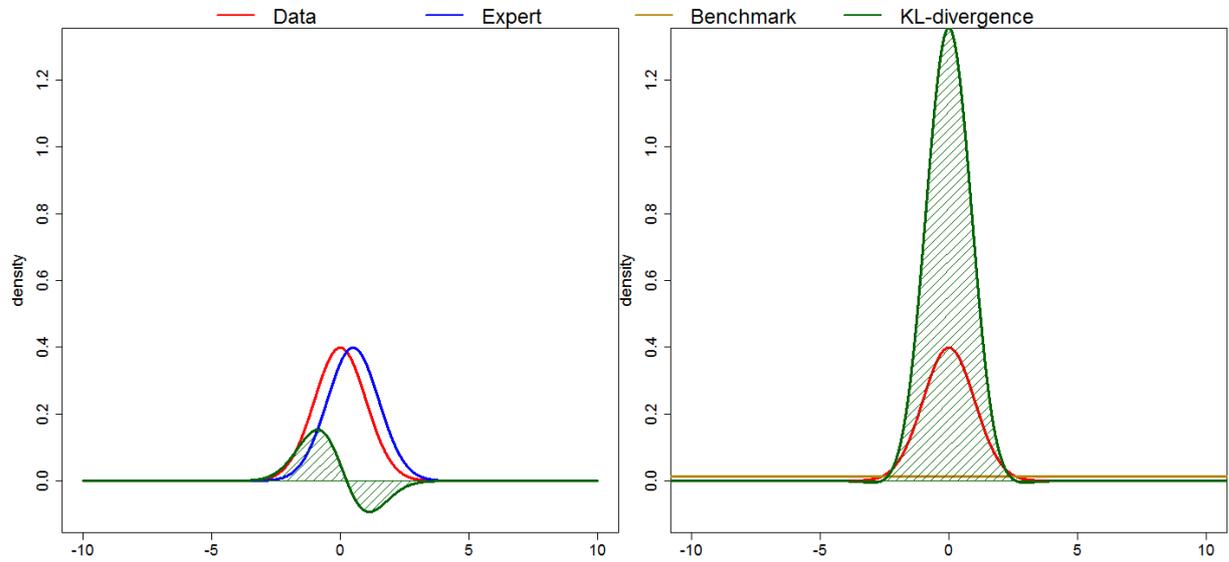

*Figure 3*. Scenarios in which there are multiple experts and one source of data. Panel A shows a experts differing in prediction and (un)certainty, all (dis)agreeing to a certain extend with the data. Panel B shows a scenario in which all experts disagree with the data, which results in the question which of the sources of information is correct?

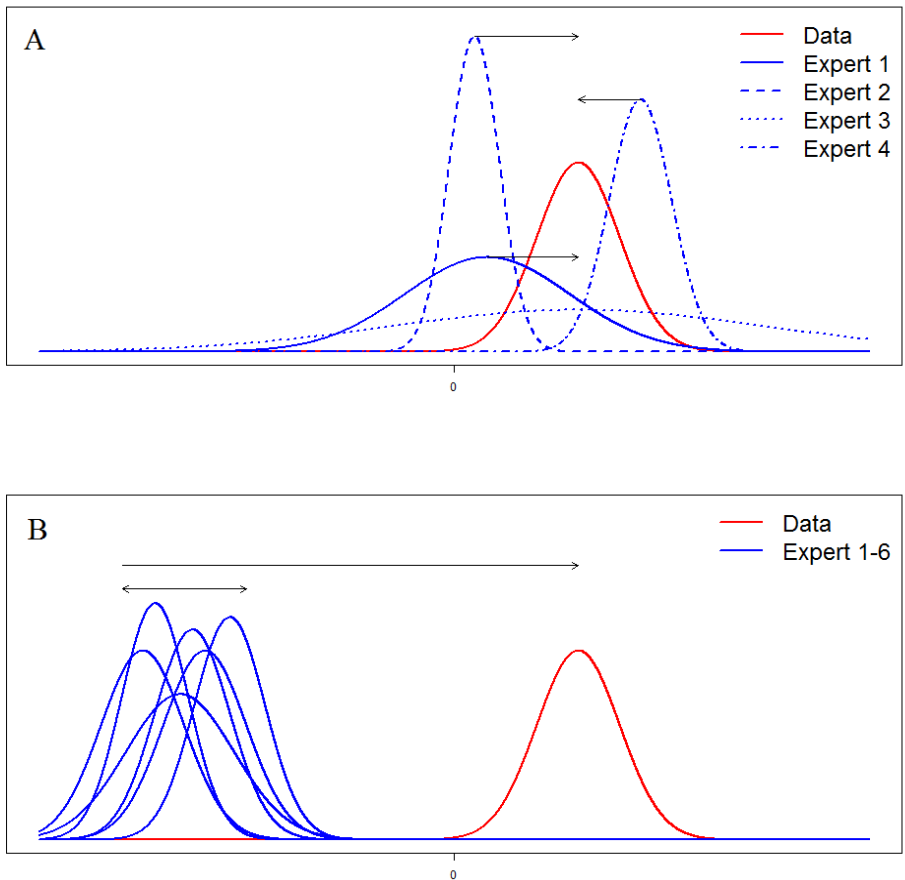

*Figure 4*. The effect on the behavior of the $DAC_d$ for different choices for benchmark priors. All panels use the same data ($N$=100) from a standard normal distribution and the same variations for $\pi_d(\theta)$ which are normal distribution for which the parameters for the mean and standard deviation are given on the x-axis and y-axis of the panels. In panel A $\pi^J(\theta) \sim N(0,10000)$, in panel B $\pi^J(\theta) \sim N(0,1)$, in panel C $\pi^J(\theta) \sim U(-50,50)$ and in panel D $\pi^J(\theta) \sim N(5,0.5)$.

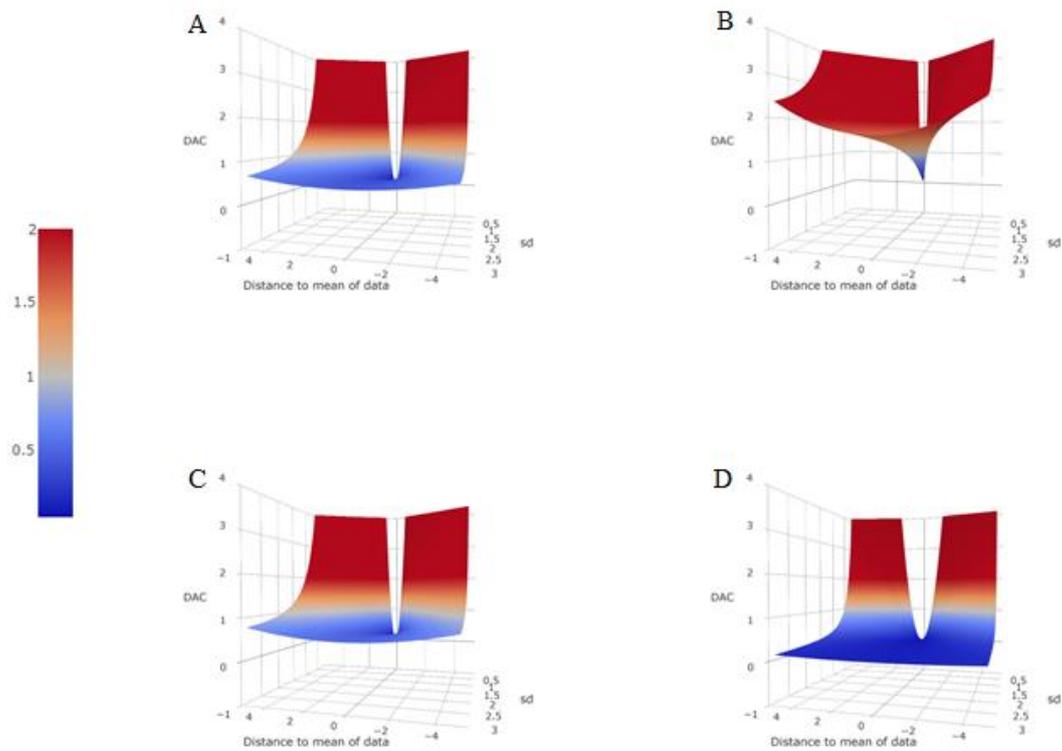

*Figure 5*. Visual presentation of all relevant distributions for the empirical study; $\pi_d(\theta)$, $\pi^J$ and $\pi^J(\theta|\mathbf{y})$.

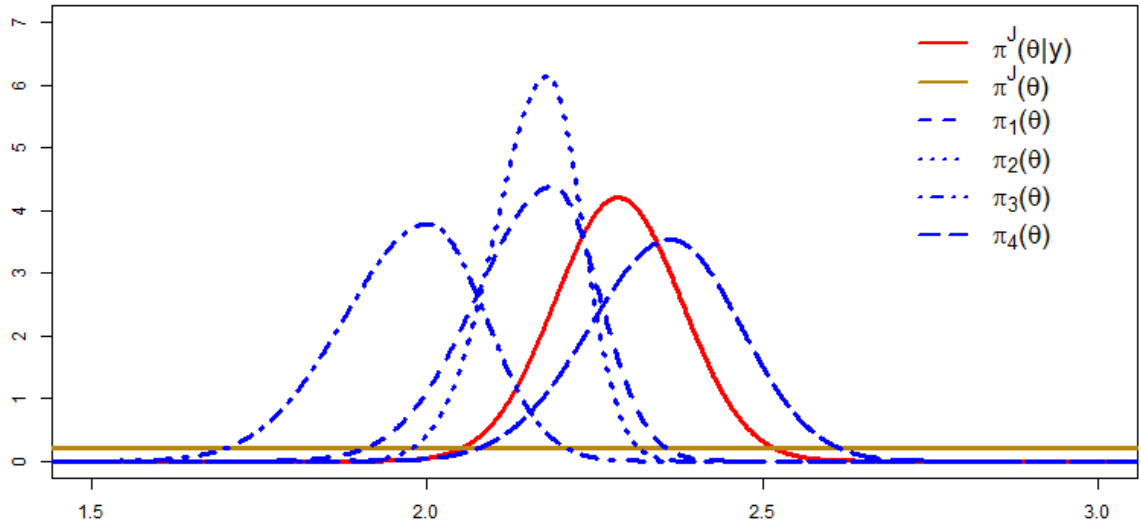

*Figure 6.* All KL divergences for $\pi_d(\theta)$ (panels A, B, C and D) and $\pi^J$ (panel E) with $\pi^J(\theta|\mathbf{y})$ as the distribution that is to be approximated. Panel A is for expert one, panel B for expert two, panel C for expert three and panel D for expert four.

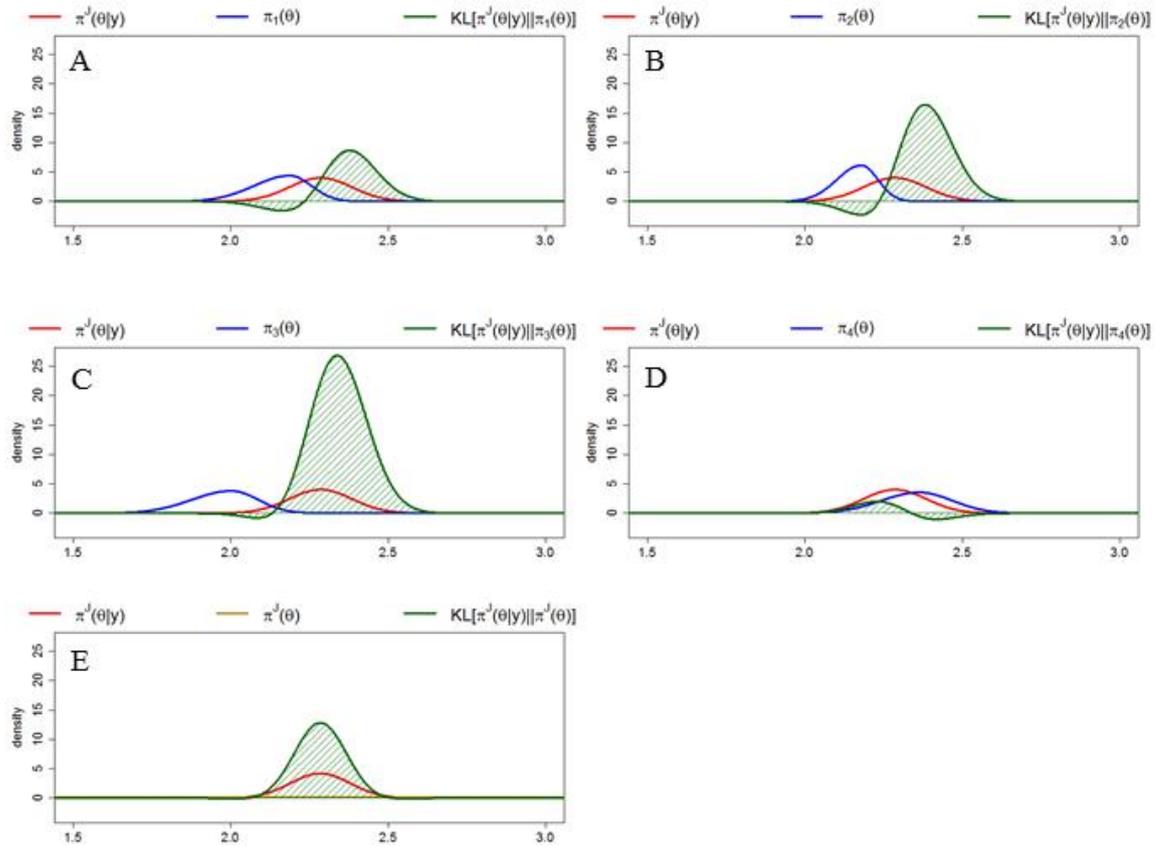

Table 1

The Values of the Hyper Parameters of $\pi_d(\theta)$ for the Empirical Study.

|          | $\mu_0$ | $\sigma_0$ | $\gamma_0$ |
|----------|---------|------------|------------|
| Expert 1 | 2.15    | 0.09       | 0.78       |
| Expert 2 | 2.16    | 0.07       | 0.82       |
| Expert 3 | 1.97    | 0.11       | 0.82       |
| Expert 4 | 2.35    | 0.11       | 0.94       |

Table 2

KL divergences for the experts' priors and the benchmark prior with $\pi^J(\theta|\mathbf{y})$ as the distribution that is to be approximated presented with resulting $DAC_d$ scores and ranking of the experts' beliefs.

|  | KL divergence | $DAC_d$ | Ranking |
|---|---|---|---|
| Expert 1 | 1.43 | 0.56 | 2 |
| Expert 2 | 2.86 | 1.12 | 3 |
| Expert 3 | 5.76 | 2.26 | 4 |
| Expert 4 | 0.19 | 0.07 | 1 |
| Benchmark | 2.55 | - | - |